\documentclass{phbauth}
\usepackage{epsf,rotate}
\newcommand{\be}{\begin{equation}}
\newcommand{\ee}{\end{equation}}
\newcommand{\ie}{{\it i.e.\ }}
\def\agt{{\buildrel >\over \sim}}

\def\vJ{{\bf J}}
\def\ve{{\bf e}}
\def\vx{{\bf x}}
\def\vy{{\bf y}}
\newcommand{\heb}{$^3$He-B}
\begin{document}
\begin{frontmatter}
\title{Magneto-acoustic rotation of transverse waves in $^3$He-B}
\author[NU]{J. A. Sauls\thanksref{thank1}},
\author[Stanford]{Y. Lee},
\author[NU]{T.M. Haard},
\author[NU]{W.P. Halperin}
\address[NU]{Department of Physics and Astronomy,
              Northwestern University, 
	      Evanston IL 60208 U.S.A.}
\address[Stanford]{present address:Department of Physics,
              Stanford University, 
	      Stanford CA 94305 U.S.A.}
\thanks[thank1]{Corresponding author. E-mail: sauls@nwu.edu} 
\begin{abstract}
In superfluid \heb,
the off-resonant coupling of the $J=2^-$, $M=\pm 1$ order parameter
collective modes to transverse current excitations
stabilizes propagating transverse waves with low 
damping for frequencies above that of the $J=2^-$ modes. 
Right- (RCP) and left circularly polarized (LCP) transverse modes
are degenerate in zero field; however, a magnetic field with 
${\bf H}||{\bf q}$ lifts this degeneracy giving rise to the
acoustic analog of circular birefringence and an acoustic Faraday
effect for linearly polarized transverse sound waves \cite{moo93}.
We present theoretical results for the temperature, pressure and
field dependence of the Faraday rotation angle,
and compare the theory with recent measurements \cite{lee98}.
The analysis provides a direct measurement 
of the Land\'e g-factor for the $J=2^-$ modes,
and new information on the magnitude of f-wave pairing
correlations in $^3$He-B.
\end{abstract}
\end{frontmatter}

The dispersion relation for RCP(+) and LCP(-) 
transverse current modes in $^3$He-B is given by \cite{moo93}
\be
\left(\frac{\omega}{q_{\pm}v_f}\right)^2 = 
\Lambda_n + \Lambda_s
\frac{\omega^2}{(\omega+i\Gamma)^2-\Omega^2_{\pm}(T,\omega,H)}
\,,
\ee
with $\Lambda_n = \frac{F^s_1}{15}\,\rho_n(\omega,T)$ and
$\Lambda_s = \frac{2F^s_1}{75}\,\rho_s(\omega,T)$.
The restoring forces are provided by quasiparticle
excitations ($\sim\rho_n$), and pair excitations
($\sim\rho_s$), with $\rho_s+\rho_n=1$.
The condensate term dominates at low temperature
($\rho_s\simeq 1$), and is
anomalously large when the sound frequency is near a resonant
frequency of the $J=2^-$, $M=\pm 1$ collective modes,
\ie when
\be
D^2_{\pm}(\omega,H,T)=(\omega + i\Gamma)^2 - 
\Omega^2_{\pm}(\omega,H,T)\,\agt\,0
\,.
\ee
The frequencies of the $M=\pm 1$ modes include the 
Zeeman splitting,
\be
\Omega^2_{\pm}=\omega_{isq}^2(T) \pm 2\,g(T)\,\omega\,\omega_L(H,T)
\,,
\ee
where $\omega_{isq}(T)$ is the frequency of the $J=2^-$ 
modes in zero field, $g(T)$ is the g-factor
for the $J=2^-$ modes and
$\omega_L(H,T)=\gamma_{\mbox{\small eff}}\,H$
is the effective
Larmor frequency that determines the linear Zeeman 
splitting of the $J=2$ multiplets \cite{hal90}.

Consider a linearly polarized transverse current excitation 
with frequency $\omega$ propagating in the $z$-direction.
The RCP and LCP modes propagate in the bulk with different 
phase velocities,
\be
\vJ(\omega,z) = 
\frac{J}{\sqrt{2}}\,\,e^{iq_+(\omega)z}\,\hat{\ve}_{+}
\,+\,
\frac{J}{\sqrt{2}}\,\,e^{iq_-(\omega)z}\,\hat{\ve}_{-}
\,,
\ee
where $\hat{\ve}_{\pm}=(\hat{\vx}\pm i\hat{\vy})/\sqrt{2}$
are the polarization vectors for RCP and LCP current modes.
The response corresponds to a pure Faraday rotation of the
polarization if the phase velocities are real, which is the
case at very low temperatures ($T\ll T_c$) and frequencies 
above the collective mode resonances 
($|\omega - \Omega_{\pm}(T)|\gg\Gamma$).
The transverse wave propagates with the average phase velocity,
$\bar{c} = 2\omega/(q_{+}+q_{-})$, while the polarization {\em rotates} with 
a spatial period, $\Lambda_H(\omega,T)=4\pi/|q_{+} - q_{-}|$.

Near the $M=+1$ mode crossing ($\omega=\Omega_{+}(T_{+})$)
the temperature dependence of the Faraday rotation period is
dominated by $q_{+}(T)\simeq q_f\,\sqrt{s(T/T_{+} -1)}$,
where $q_f=\omega/v_f$ and
$s\equiv 2|\Omega_{+}^{'}|T_{+}/\Lambda_s\Omega_{+}$
are evaluated at $T=T_{+}$.
The field dependence of $\Lambda_H$
originates from the Zeeman splitting of the $M=\pm 1$
modes, and appears through the shift in the wavenumber 
of the LCP mode. Using
$D_{-}^2 = D_{+}^2 +4\,g(T)\,\omega\,\omega_L(T,H)$
we can write, $q_{-}\simeq q_f\sqrt{s(T/T_{+}-1)+B/B_{+}}$,
where we set $g=g(T_{+})$ for low fields,
$g\omega_L\ll\Omega_{+}$, and $T\,\agt\,T_{+}$. 
The formula for $q_{-}(T)$ also simplifies 
because $\Lambda_n D_{-}^2(T) \ll \Lambda_s\omega^2$.
The field dependence is determined by the field scale,
$B_{+}\equiv\Lambda_s\Omega_{+}/
4\,g\,\gamma_{\mbox{\small eff}}$, where
$\gamma_{\mbox{\small eff}}$,
is the effective gyromagnetic ratio.
Thus, near the mode crossing the Faraday rotation period 
simplifies to
\be\label{Faraday_Period}
\Lambda_H=\frac{\lambda_f}{\sqrt{\alpha+\beta}-\sqrt{\alpha}}
\,,
\ee
where $\lambda_f=4\pi v_f/\Omega_{+}$,
$\alpha = s\,(T/T_{+} -1)$, and
$\beta = H/B_{+}$. Except for temperatures very close to
the collective mode resonance or for very low fields;
i.e. for $s(T/T_{+}-1)>H/B_{+}$, $\Lambda_H$
scales inversely with $H$ and  as the square
root of the reduced temperature,
$\Lambda_H\simeq K\sqrt{T/T_{+}-1}/H$,
where $K\simeq\xi_{\Omega}
(\pi v_f/\gamma_{\mbox{\small eff}} g)
\sqrt{\frac{2}{25}(m^*/m-1)\rho_s/\rho}$ 
is determined by the effective mass, superfluid
density, gyromagnetic ratio, Land\'e
g-factor for the modes, and the slope of the collective
mode frequency, which is approximately
$\xi_{\Omega}=8|\Omega_{+}'|T_{+}/\Omega_{+}\simeq 1$
at $T\simeq 0.44 T_c$. Recent experimental measurements 
of the Faraday rotation angle \cite{lee98} 
confirm this scaling behavior.
A quantitative comparison between the theory and experimental
measurements of the Faraday rotation period as a function of
temperature, pressure and field can be made on the basis
of Eq. \ref{Faraday_Period}.

\begin{figure}[ht]
\centerline{\epsfysize=\hsize\rotate[r]{{\epsfbox{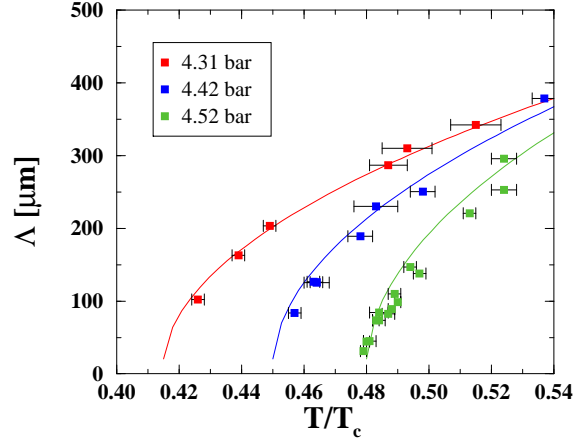}}}}
\caption{\small Comparison between the theoretical and
experimental results for the temperature
dependence of the Faraday rotation period for a field
of $H=100\,\mbox{Gauss}$, frequency of 
$\omega/2\pi=82.26\,\mbox{MHz}$
at pressures, $p=4.31\,\mbox{bar}$
(red), $p=4.42\,\mbox{bar}$ (blue), and $p=4.52\,\mbox{bar}$ (green)
\cite{lee98}.}
\label{Fig:Faraday_Period}
\end{figure}

Fig. \ref{Fig:Faraday_Period} 
shows the comparison between the theoretical and
experimental results for the temperature dependence of the Faraday
rotation period scaled to $H=100\,\mbox{Gauss}$
for three pressures near $4\,\mbox{bar}$.
The inputs to the theoretical calculation are: the
Fermi liquid parameters, $T_c$, $v_f$, $F_1^s$, $F_0^a$
\cite{hal90}, the mode data, $T_{+}/T_c$ and $\omega$,
and the gyromagnetic ratio for $^3$He.
The calculated parameters that enter Eq.
\ref{Faraday_Period} are the effective Larmor frequency, 
$\omega_{L}=\gamma_{\mbox{\small eff}}H$, the condensate response, 
$\rho_s(\omega,T_{+})$ and the slope of the mode frequency,
$|\Omega_{+}^{'}|T_{+}$. 
The remaining open parameter is the Land\'e g-factor which
determines the vertical scale for $\Lambda_H$; the 
fit to the data gives $g=0.02\pm 0.002$, which is in 
agreement with theoretical calculations of the g-factor that
include attractive f-wave pairing 
correlations in \heb \cite{sau82b}.

\end{document}